\documentclass[11pt,a4paper]{article}

\usepackage[margin=2.5cm]{geometry}
\usepackage[T1]{fontenc}
\usepackage{lmodern}
\usepackage{amsmath,amssymb,bm}
\usepackage{microtype}
\usepackage{booktabs}
\usepackage{cite}
\usepackage{graphicx}
\usepackage[hidelinks,pdfencoding=auto]{hyperref}
\hypersetup{
  pdftitle={Charged spin-3/2 in a chiral Feshbach--Villars model},
  pdfauthor={Haryanto M. Siahaan},
  pdfsubject={Hyperbolicity, Landau-level instability, and exceptional points},
  pdfkeywords={spin 3/2, Feshbach--Villars, pseudo-Hermiticity, Landau levels, exceptional points}
}
\newcommand{\vS}{\vec S}\newcommand{\vB}{\vec B}\newcommand{\vE}{\vec E}
\newcommand{\vp}{\vec p}\newcommand{\vA}{\vec A}\newcommand{\vpi}{\vec\pi}
\newcommand{\tp}{(\tau_3+i\tau_2)}
\newcommand{\dla}{\langle\!\langle}\newcommand{\dra}{\rangle\!\rangle}
\newcommand{\eq}[1]{Eq.~(\ref{#1})}
\newcommand{\sect}[1]{Sec.~\ref{#1}}
\newcommand{\fig}[1]{Fig.~\ref{#1}}
\newcommand{\tabl}[1]{Table~\ref{#1}}

\begin{document}
\title{Charged spin-$\frac32$ in a chiral Feshbach--Villars model:\\
hyperbolic propagation, Landau-level instability,\\ and exceptional points}

\author{Haryanto M. Siahaan\\[4pt]
\small\itshape Program Studi Fisika, Universitas Katolik Parahyangan,\\
\small\itshape Jalan Ciumbuleuit 94, Bandung 40141, Indonesia\\[2pt]
\small\texttt{haryanto.siahaan@unpar.ac.id}}

\date{}

\maketitle

\begin{abstract}
We study the Pauli-coupled second-order wave equation on
$(\frac32,0)\oplus(0,\frac32)$ with a free gyromagnetic ratio $g$, an equation known from the
arbitrary-spin literature and obtained here from the Poincar\'e-projector construction.
Its Feshbach--Villars form is a sixteen-component pseudo-Hermitian Hamiltonian
system; the covariant characteristics are lightlike in smooth electromagnetic backgrounds, as known.
The solution space is chirality doubled; a covariant physical single-species
reduction is not assumed. The exact spin-$j$ Landau spectrum remains real
at every field strength precisely when $|g|j\le1$, the complement of the known condition for
magnetic pair production. At $g=2$, the critical spin-$\frac32$ channel reaches zero at
$|qB|=m^2/2$, where each chiral block develops a second-order exceptional point.
This threshold also marks the loss of a time-independent positive-definite conserved
quadratic form for the channel. We also include a bounded, Gaussian-damped harmonic
electrostatic potential and its associated electric field. The complete first-order
problem on a degenerate manifold reduces to $V\pm iC$, with $V$ and $C$ the
projected potential and spin--electric coupling. On the lowest two-state
spin-$\frac32$ manifold, nonreal eigenvalue branches arise for
$m^2/6<|qB|<m^2/2$ at sufficiently small nonzero perturbation amplitude, except
at a special well range where the first-order splitting vanishes. The spin-$\frac12$,
$g=2$ limit satisfies the Dirac spectral-reality check, whereas $g=1/j$ has no
imaginary first-order correction for this perturbation. These results separate spectral
instability from loss of hyperbolicity; they do not establish a physical fermionic quantization.
\end{abstract}

\medskip
\noindent\textbf{Keywords:} higher-spin wave equations; Feshbach--Villars transformation;
pseudo-Hermiticity; Landau levels; exceptional points; spectral instability.

\section{Introduction}\label{sec:intro}

A consistent local description of a massive charged spin-$\frac32$ particle in an external
electromagnetic field remains a demanding problem. In the Rarita--Schwinger (RS)
formulation~\cite{FierzPauli1939,RaritaSchwinger1941}, the vector-spinor $\psi_\mu$
contains spin-$\frac12$ sectors that must be removed by constraints. Minimal electromagnetic
coupling modifies these constraints and leads to several distinct difficulties. The
quantized theory encounters a failure of Lorentz covariance~\cite{JohnsonSudarshan1961},
while the classical characteristics become superluminal in arbitrarily weak nonzero
magnetic fields. Hyperbolicity is lost at $|eB|=\frac32m^2$~\cite{VeloZwanziger1969a,VeloZwanziger1969b}.
At the same field strength, the Landau energies become complex and the norm becomes
indefinite~\cite{Seetharaman1975}. Generic nonminimal couplings do not remove these
problems~\cite{DeserPascalutsaWaldron2000}. Consistent constructions in constant backgrounds
instead require a specific set of such couplings~\cite{PorratiRahman2009}, with string theory
providing a related realization together with the gyromagnetic ratio
$g=2$~\cite{PorratiRahmanSagnotti2011,FerraraPorratiTelegdi1992}.

An alternative approach starts from a second-order equation rather than a constrained
first-order one. Napsuciale, Kirchbach, and Rodr\'{\i}guez (NKR)~\cite{NKR2006}
use the Poincar\'e Casimirs $p^2$ and $\mathcal W^2$ to construct a covariant
mass-$m$, spin-$\frac32$ projector on $\psi_\mu$ and the corresponding wave equation.
Its minimal gauging gives $g=\frac13$ and is acausal, whereas causal propagation
selects $g=2$, together with a $g=0$ branch. The value $g=2$ is also favored by the
high-energy behavior of Compton scattering~\cite{Weinberg1970,FerraraPorratiTelegdi1992,DelgadoAcostaNapsuciale2009}.
For spin $\frac12$, the analogous second-order description is the
Feynman--Gell-Mann equation~\cite{FeynmanGellMann1958}. Its linearization in time by
the Feshbach--Villars (FV) procedure~\cite{FeshbachVillars1958} leads to the
eight-component equation of Robson and Staudte~\cite{RobsonStaudte1996,Staudte1996}.
The associated indefinite-metric perturbation theory has recently been shown to reproduce
the Dirac Stark shifts of hydrogenic ions through second order~\cite{TST2025}.
Hamiltonian formulations on $(j,0)\oplus(0,j)$ have a longer history, beginning with
Weaver, Hammer, and Good~\cite{WeaverHammerGood1964}, Mathews~\cite{Mathews1966},
and Guertin~\cite{Guertin1974}. More recent second-order developments by Delgado Acosta,
Kirchbach, Napsuciale, and collaborators~\cite{DelgadoAcosta2011,DelgadoAcosta2012,Acosta2015}
include a pure spin-$\frac32$ construction with Belinfante's value $g=\frac23$~\cite{Acosta2015}.
The second-order equation on $(s,0)\oplus(0,s)$ itself is older. For the normal magnetic
moment, $g=1/s$, it goes back to Hurley~\cite{Hurley1971,Hurley1972,Hurley1974}.
Kruglov~\cite{Kruglov2001} extended it to an arbitrary gyromagnetic ratio as a generalization
of the Feynman--Gell-Mann equation, noted that its characteristic surfaces remain lightlike in
external fields, and used its exact solutions in constant uniform fields to obtain the
pair-production probability and the one-loop effective Lagrangian for arbitrary spin. In agreement
with the quasiclassical analysis of Marinov and Popov~\cite{MarinovPopov1972}, pairs are
produced in a purely magnetic field when $gs>1$.

The present work brings these two approaches together in the chiral representation
$(\frac32,0)\oplus(0,\frac32)$. In this representation, the Poincar\'e content is
pure spin $\frac32$, so the spin projector is the identity and the free equation is
Klein--Gordon on each of the four components of either chirality. There are consequently
no lower-spin sectors or differential constraints to eliminate. Within the class of
kinetic tensors specified below, gauging introduces a single parity-even Pauli term
with a coefficient $g$ that remains free. The resulting equation coincides with the
arbitrary-spin equation of Refs.~\cite{Hurley1971,Hurley1972,Hurley1974,Kruglov2001}; the projector route explains
why it needs no constraints and how it is related to the vector-spinor theory. Applying the FV procedure then gives eight
components per chirality, or sixteen in total, equipped with the indefinite metric
used in Refs.~\cite{Staudte1996,TST2025}.

The equation, its causal propagation, its spectrum in constant fields, and the magnetic
threshold $gs>1$ are therefore not new. The contribution of the present paper is the
Feshbach--Villars Hamiltonian formulation of this equation and what that formulation
exposes: the pseudo-Hermitian structure and the counting of states, the exceptional-point
form of the magnetic threshold together with the loss of a positive conserved form, the
mixed-norm degenerate perturbation theory for electric fields, and a bounded electrostatic
example with nonreal eigenvalue branches below the magnetic threshold.

This construction also clarifies the relation to the first-order proposal discussed
in Ref.~\cite{Siahaan2007}. The customary covariant relation between the two chiralities
of $(j,0)\oplus(0,j)$ has momentum order $2j$~\cite{Weinberg1964}; an eight-component
Dirac-like equation obtained simply from the spin-$\frac32$ matrices therefore cannot
square to the Klein--Gordon operator. The FV equation considered here is instead first
order only in time, quadratic in spatial momentum, and pseudo-Hermitian. It should
accordingly be understood as a time-linearized chiral second-order model, rather than
as an established rewriting of the interacting RS theory.

The paper is organized as follows. Section~\ref{sec:fv} introduces the equation,
performs the FV transformation, and explains the enlarged state space. It also develops
the mixed-norm electric perturbation problem and a bounded-background example in which
nonreal eigenvalue branches arise. Section~\ref{sec:hyp} examines the characteristics,
and Sec.~\ref{sec:landau} derives the exact magnetic spectrum and its channel exceptional
point. Section~\ref{sec:compare} distinguishes these results from RS propagation
pathologies and discusses their effective-theory interpretation. The conclusions and
remaining questions are collected in Sec.~\ref{sec:outlook}. Throughout, the term
``model'' refers to the specified wave equation~\eqref{eq:2nd}.

\section{The chiral second-order equation and its Feshbach--Villars form}\label{sec:fv}

\subsection{Representation and equation}\label{sec:eq}

We use units $\hbar=c=1$, the metric $(+,-,-,-)$, real charge $q$ and gyromagnetic ratio $g$, and mass $m>0$.
With $D_\mu=\partial_\mu+iqA_\mu$, we have
$[D_\mu,D_\nu]=iqF_{\mu\nu}$, and $F_{0i}=E_i$, $F_{ij}=-\epsilon_{ijk}B_k$. On the chiral representation $(j,0)$ the
Lorentz generators are $S^{ij}=\epsilon_{ijk}S_k$, $S^{0i}=K_i=-iS_i$, with $\vS$ the spin-$j$ matrices; on $(0,j)$,
$K_i=+iS_i$. With these conventions, contraction with the electromagnetic field tensor gives
\begin{equation}\label{eq:SF}
\tfrac12S^{\mu\nu}F_{\mu\nu}=-\vS\!\cdot\!(\vB\pm i\vE),\qquad\text{upper sign for }(j,0),
\end{equation}
and the generators are self-dual, $\frac12\epsilon^{\mu\nu\rho\sigma}S_{\rho\sigma}=\pm iS^{\mu\nu}$.

The Poincar\'e-projector equation of NKR, $[-\Gamma_{AB\mu\nu}p^\mu p^\nu+m^2\delta_{AB}]\Psi^B=0$ with
$\Gamma_{AB\mu\nu}=-\frac{1}{2s}(T_{AB\mu\nu}+s(s-1)\delta_{AB}g_{\mu\nu})$ and $\mathcal W^2=T_{AB\mu\nu}p^\mu p^\nu$
[Eqs.~(16), (23), (24) of Ref.~\cite{NKR2006}], takes a particularly simple form on $(\frac32,0)\oplus(0,\frac32)$. Since this
representation contains no spin other than $\frac32$, we have $\mathcal W^2=-\frac{15}{4}p^2$ identically, and
$\Gamma_{AB\mu\nu}=g_{\mu\nu}\delta_{AB}$ up to a part antisymmetric in $(\mu\nu)$, which commuting momenta do not see.
The free equation therefore reduces to $(p^2-m^2)\Xi=0$ componentwise. This observation also identifies the freedom that remains when the equation is gauged: the free theory fixes the symmetric part of the kinetic tensor, whereas its antisymmetric part generates the coupling to $F_{\mu\nu}$. If, as in Ref.~\cite{NKR2006}, the kinetic tensor is built from the
representation matrices alone---no further dimensionful coefficients, no derivatives of $F_{\mu\nu}$---the antisymmetric part,
which in $\psi_\mu$ is the five-parameter family of Eq.~(120) of Ref.~\cite{NKR2006}, reduces on $(j,0)\oplus(0,j)$ to a single
parity-even term, $\Gamma^{\mu\nu}=g^{\mu\nu}-i\,g\,S^{\mu\nu}$. The counting is representation-theoretic.
Matrices acting within one chiral block transform as $(j,0)\otimes(j,0)=\bigoplus_{k=0}^{2j}(k,0)$, which contains
$(1,0)$ exactly once, and matrices connecting the two blocks transform as $(j,0)\otimes(0,j)=(j,j)$, which contains
neither $(1,0)$ nor $(0,1)$. The antisymmetric rank-two covariants on $(j,0)\oplus(0,j)$ are therefore spanned by
$S^{\mu\nu}$ and its dual, $\frac12\epsilon^{\mu\nu\rho\sigma}S_{\rho\sigma}=i\rho_3S^{\mu\nu}$, where
$\rho_3={\rm diag}(1,-1)$ in the chirality space. On the full representation the dual is linearly independent of
$S^{\mu\nu}$; what excludes it is parity, since its coupling to $F_{\mu\nu}$ is an electric-dipole term.
Under these assumptions, the interacting chiral equation can be written as
\begin{equation}\label{eq:2nd}
\Bigl[D_\mu D^\mu+m^2+\frac{g\,q}{2}\,S^{\mu\nu}F_{\mu\nu}\Bigr]\Xi_{L,R}=0,\qquad
\frac{g\,q}{2}\,S^{\mu\nu}F_{\mu\nu}=-g\,q\,\vS\!\cdot\!(\vB\pm i\vE),
\end{equation}
with $\Xi_L\in(j,0)$, $\Xi_R\in(0,j)$, $2j+1$ components each. For $j=\frac12$, $S^{\mu\nu}=\frac12\sigma^{\mu\nu}$ and
\eq{eq:2nd} with $g=2$ is exactly $(i\gamma^\mu D_\mu+m)(i\gamma^\nu D_\nu-m)\psi=-(D^2+m^2+\frac q2\sigma^{\mu\nu}F_{\mu\nu})\psi$
written in the Weyl basis, i.e.\ the Feynman--Gell-Mann equation~\cite{FeynmanGellMann1958} and the second-order
equation underlying the eight-component theory~\cite{RobsonStaudte1996,TST2025}.
For general spin, \eq{eq:2nd} is the equation postulated in Ref.~\cite{Kruglov2001}
[Eq.~(1) there, with gyromagnetic ratio $q/s$ in that notation], which reduces at $g=1/s$ to
the equation of Hurley~\cite{Hurley1971,Hurley1972,Hurley1974}. The derivation above adds the
statement that, on this representation, it is the only equation of the stated class.

Before applying the FV transformation, it is useful to clarify the scope of
\eq{eq:2nd}. Within the stated class of parity-even kinetic tensors, $g$ is the only
remaining coupling parameter, but the construction itself does not determine its value.
We therefore retain $g$ and use two familiar values as benchmarks. The first is $g=2$,
which is selected by the high-energy arguments of
Refs.~\cite{Weinberg1970,FerraraPorratiTelegdi1992}, by causality in the vector-spinor
second-order construction~\cite{NKR2006}, and in the string-theory setting of
Ref.~\cite{PorratiRahmanSagnotti2011}. The second is Belinfante's value $g=1/j$,
which gives $g=\frac23$ for the minimally coupled RS equation and for the pure
spin-$\frac32$ construction of Ref.~\cite{Acosta2015}. Whether scattering consistency
selects either value in the present model, or requires additional structure, is not
established here. This question is revisited in Sec.~\ref{sec:outlook}.

Gauge invariance alone does not remove this freedom. In the second-order formulation,
the one-photon and seagull vertices derived from the same gauged kinetic tensor satisfy
the Compton Ward identity for arbitrary $g$. The corresponding algebra is displayed
for the vector-spinor theory in Ref.~\cite{SiahaanThesis}. This separates the requirement
of gauge invariance from the further restrictions that may follow from a scattering
or physical-state analysis.

The restriction to the stated kinetic tensor is also important for interpreting strong-field
results. Additional Lorentz- and gauge-invariant operators can be written, for example
$(c\,q^2/\Lambda^2)F_{\mu\nu}F^{\mu\nu}$ or terms involving derivatives of $F_{\mu\nu}$.
The coefficient $\Lambda^{-2}$ may or may not introduce a scale different from $m$.
Such terms can leave the principal symbol and weak-field magnetic moment unchanged
while modifying the spectrum at $|qB|\sim m\Lambda$. They are excluded here by the
operator truncation that defines the model, not by dimensional analysis. The strong-field
results below are exact for \eq{eq:2nd}; their interpretation within an effective theory
requires the additional hierarchy discussed in Sec.~\ref{sec:compare}.

\subsection{Feshbach--Villars linearization}\label{sec:fvlin}

To express the second-order equation as an evolution equation, let us introduce
$\pi_0=i\partial_t-qA_0$ and $\vpi=\vp-q\vA$. These definitions give
$D_\mu D^\mu=-\pi_0^2+\vpi^{\,2}$, so it is convenient to define, for each chirality,
\begin{equation}\label{eq:K}
K_{L,R}=\vpi^{\,2}-g\,q\,\vS\!\cdot\!(\vB\pm i\vE),\qquad \pi_0^2\,\Xi=(K+m^2)\,\Xi .
\end{equation}
The FV transformation separates the wavefunction and its first covariant time derivative
into the two fields $\varphi=\frac1{2m}(m+\pi_0)\Xi$ and
$\chi=\frac1{2m}(m-\pi_0)\Xi$. In particular,
\[
\Xi=\varphi+\chi,\qquad \pi_0\Xi=m(\varphi-\chi).
\]
Substituting these relations into \eq{eq:K} converts the second-order equation into a
first-order system for $\Psi=(\varphi,\chi)^T$. The resulting equation is
$i\partial_t\Psi=H\Psi$, with
\begin{equation}\label{eq:HFV}
H_{L,R}=\tp\otimes\frac{K_{L,R}}{2m}+m\,\tau_3\otimes\mathbf 1_{2j+1}+qA_0\,\mathbf 1 ,
\end{equation}
where the Pauli matrices $\tau_i$ act on the FV doublet and the combination $\tp$ is nilpotent.
The doubling takes place separately within each chirality. Consequently, each chiral block
has $2(2j+1)$ components, eight when $j=\frac32$, and the full system is
$H={\rm diag}(H_L,H_R)$ on $4(2j+1)$ components. Writing $\rho_i$ for Pauli matrices in the chirality space
$(L,R)$, the conserved indefinite inner product is
\begin{equation}\label{eq:tau5}
\dla\Psi|\Phi\dra=\int\Psi^\dagger\tau_5\Phi\,d^3r,\qquad\tau_5=\rho_1\otimes\tau_3\otimes\mathbf 1_{2j+1},
\end{equation}
which for $j=\frac12$ is the metric $\tau_3\otimes\gamma^0$ of Refs.~\cite{Staudte1996,TST2025} in the Weyl basis; $\rho_1$
is the matrix that represents $\gamma^0$, i.e.\ the parity-invariant pairing of $(j,0)$ with $(0,j)$.

The conservation of this inner product follows from the pseudo-Hermiticity of the
Hamiltonian. To verify this property, we use the Hermiticity of the spin matrices,
$\vS^\dagger=\vS$, together with the reality of the prescribed fields $\vE$ and $\vB$.
Combining these properties with $\tau_3(\tau_3-i\tau_2)\tau_3=\tau_3+i\tau_2$ gives, for arbitrary external fields,
\begin{equation}\label{eq:pseudo}
\tau_3H_R^\dagger\tau_3=H_L,\qquad \tau_5H^\dagger\tau_5=H .
\end{equation}
For nonzero $g$, an electric spin coupling generally prevents either block from being $\tau_3$-pseudo-Hermitian by
itself; when $\vE=0$, $H_L=H_R$. These operator identities are understood on compatible domains with boundary terms
vanishing. Under these assumptions, $H^\dagger\tau_5=\tau_5H$ ensures that
$\dla\Psi|\Phi\dra$ is conserved for solutions of the evolution equation. The calculation
is the spin-$j$ counterpart of that in Ref.~\cite{TST2025}, with $\vec\sigma/2$ replaced by $\vS$.

The free limit provides a direct check of the construction. Using the nilpotent identity
$\tp^2=0$ together with $\{\tau_3,\tau_3+i\tau_2\}=2$, the square of the
Hamiltonian reduces to a scalar operator. It follows that
\begin{equation}\label{eq:free}
\begin{gathered}
H_L^2=(\vp^{\,2}+m^2)\mathbf 1,\qquad
\det\bigl[k_0-H(\vec k)\bigr]=(k_0^2-\vec k^2-m^2)^{2(2j+1)},\\
G(p)=\frac{p_0+H(\vp)}{p_0^2-\vp^{\,2}-m^2+i\varepsilon}.
\end{gathered}
\end{equation}
The full Hamiltonian therefore has eigenvalues $\pm E_p$, each with multiplicity
$2(2j+1)$. The same algebra will be useful in an external field: whenever $K$ reduces
to a $c$-number in a fixed channel,
$[\tp\frac{K}{2m}+m\tau_3]^2=(K+m^2)\mathbf 1$; this is used in \sect{sec:landau}. The propagator has a single mass shell, with the two energy poles $p_0=\pm E_p$, and no
lower-spin pieces, in contrast with the vector-spinor propagator of Ref.~\cite{NKR2006}, whose off-shell spin-$\frac12$
sector carries a gauge-parameter dependence [their Eqs.~(54), (80), (102)--(108)]; it does, however, propagate both
chirality sectors, which is the subject of the next subsection.

The simplicity of this time linearization is specific to the chiral second-order equation.
In the vector-spinor construction, by contrast, the coefficient of $p_0^2$
in the NKR kinetic
tensor $\Gamma_{\alpha\beta\mu\nu}$ [Eq.~(38) of Ref.~\cite{NKR2006}] is a $16\times16$ matrix $\Gamma_{\alpha\beta00}$ of
rank eight, with the $\psi_0$ rows and columns identically zero (a consequence of $p^\alpha K_{\alpha\beta}=0$, their Eq.~(39)).
Only half the components carry a second time derivative; a Hamiltonian form requires eliminating the other half through the
gauged constraints [their Eqs.~(126)--(127)], which is where $F_{\mu\nu}$ enters at first-derivative order and where the
$g$-dependence of causality originates. In \eq{eq:2nd} the coefficient of $\partial_t^2$ is $\mathbf 1_{2j+1}$, so the FV
form is immediate. This is also the relation to the Hamiltonians of Refs.~\cite{WeaverHammerGood1964,Mathews1966,Guertin1974}:
those linearize Weinberg's $2j$-order equation in time without enlarging the space and are of degree $2j$ in $\vp$;
\eq{eq:HFV} is quadratic because the FV doubling absorbs the rest. At $j=\frac12$ both describe the Dirac particle once the
first-order relation between the chiralities is imposed; the unrestricted squared system remains doubled, as we now discuss.
A different first-order form of \eq{eq:2nd} was constructed in Ref.~\cite{Kruglov2001}: there the wavefunction is
supplemented by all four covariant derivatives, $(\Xi,-D_\mu\Xi/m)$, which gives a covariant
Duffin--Kemmer-type system with $5(2s+1)$ components per chirality, suited to covariant
calculations with density matrices. The FV form uses the time derivative only; it is not
manifestly covariant, but it is a Hamiltonian, and the spectral and norm questions studied
below are posed in it.

\subsection{What the sixteen components describe, and what the equation does not fix}\label{sec:states}

The absence of lower-spin components does not by itself settle the number of physical
particle species. It is therefore useful to examine the free state count before turning
to interactions. At rest, $H_0=m\,\mathbf 1_\rho\otimes\tau_3\otimes\mathbf 1$, and the
positive-energy projector $P_+=\frac12(1+H_0/m)$ has rank $2(2j+1)$, eight for $j=\frac32$. The system thus has
$2(2j+1)$ positive- and $2(2j+1)$ negative-frequency modes, twice the content of one spin-$j$ particle and its
antiparticle; and on the positive-energy subspace the metric \eqref{eq:tau5} restricts to $\rho_1\otimes\mathbf 1$, of
signature $(2j{+}1,\,2j{+}1)$. The positive-energy sector itself carries both norm signs. This is not a new feature of
spin $\frac32$: for $j=\frac12$ the same count gives the known structure of the eight-component theory, whose
positive-energy sector has signature $(2,2)$, the negative-norm half being the solutions of
$(i\gamma^\mu D_\mu+m)\psi=0$, the mass $-m$ copy~\cite{RobsonStaudte1996,Staudte1996,TST2025}. There the physical
sector is selected by the first-order Dirac relation $\psi_R=(i\bar\sigma^\mu D_\mu/m)\psi_L$, which is gauge covariant
and preserved by the dynamics because the second-order operator factorizes,
$(i\sigma^\mu D_\mu)(i\bar\sigma^\nu D_\nu)=-(D^2+\frac q2\sigma^{\mu\nu}F_{\mu\nu})$. At rest this relation reads
$\rho_1\Xi={\rm sgn}(E)\,\Xi$: the Dirac solutions are the $\tau_5=+1$ states, of positive norm for either sign of the
energy, which is the content of the Robson--Staudte identification of the physical solutions with the positive-norm
ones.

The same doubling can be followed explicitly in a magnetic background. Let us consider
a time-independent purely magnetic field in a gauge with $A_0=0$.
Then $H_L=H_R$ and $H=\mathbf 1_\rho\otimes H_L$ acts trivially on the chirality index.
In the real, gapped regime, a normalized eigenvector $u$ of $H_L$ with energy $E$ has FV charge
$\int u^\dagger\tau_3u\,d^3r=\nu={\rm sgn}\,E$. Its two-dimensional chirality-partner space
$\{(a u,b u)\}$ has Gram matrix
$\nu\bigl(\begin{smallmatrix}0&1\\1&0\end{smallmatrix}\bigr)$ and signature $(1,1)$.
Additional orbital or spin degeneracies give further such pairs. Since $[H,\rho_1]=0$,
the subspaces $\Psi_R=\pm\Psi_L$ are invariant, and the metric restricts, after a constant normalization,
to $\pm\tau_3\otimes\mathbf 1$. Before a spin channel is fixed, each reduction has $2j+1$
positive- and $2j+1$ negative-frequency polarizations per orbital mode, with the conserved FV charge form.
The norm sign of a stable mode is $\eta s$, where $\rho_1=\eta$ and $s={\rm sgn}\,E$.
The magnetic channel results below apply to either reduction.

Although these reductions have the expected mode count, invariance under time evolution
is not sufficient to identify a physical theory. More generally, any fixed operator $M$ for which
$\Psi_R=M\Psi_L$ is well defined and $[M,H_L]=0$ defines an invariant graph subspace.
Maximal positive subspaces of the chirality-partner space are not unique: in the real, gapped regime,
$\Psi_R={\rm sgn}(H_L)\Psi_L$ is an invariant choice with positive $\tau_5$-norm for either frequency sign.
A physical single-species interpretation additionally requires an appropriate covariant relation between the
chiralities. For $j=\frac12$ the Dirac relation provides such a choice and is preserved by the interaction;
at rest it selects $\rho_1={\rm sgn}(E)$, not $\rho_1=+1$ for both branches.
For $j=\frac32$ the customary free Weinberg--Joos relation is of order $2j=3$ in momentum~\cite{Weinberg1964}.
Its preservation after gauging has not been established for the present model.
Thus the magnetic-frame reductions have a single spin-$j$ sector's mode count and charge form, but their
identification with a covariant physical single-species theory is not established here.

\subsection{Degenerate electric perturbations}\label{sec:electric}

We now use this state-space structure to formulate perturbation theory for an electric
field. The part of the Hamiltonian that acts directly on the spin is
\begin{equation}\label{eq:electricoperator}
\begin{aligned}
\delta H_E&=-\frac{igq}{2m}\,\rho_3\otimes\tp\otimes\vS\!\cdot\!\vE,\\
\tau_5\delta H_E&=-\frac{gq}{2m}\,\rho_2\otimes(1+\tau_1)\otimes\vS\!\cdot\!\vE.
\end{aligned}
\end{equation}
The second line is Hermitian and off-diagonal in the eigenbasis of $\rho_1$.
Its diagonal matrix elements therefore vanish in a state of definite chirality-exchange parity.
This does not determine a first-order energy shift: the opposite-signature partners are degenerate, and a
static electric field must be included together with its potential, $\vE=-\nabla A_0$.

To include both effects consistently, let $\{u_a\}_{a=1}^d$ span an isolated positive-frequency eigenspace of the unperturbed magnetic operator,
with common energy $\mu$ and unit FV charge. Write its second-order wavefunctions as
$\Xi_a=\varphi_a+\chi_a=\sqrt{m/\mu}\,f_a$, where
\begin{equation}\label{eq:normalization}
\int f_a^\dagger f_b\,d^3r=\delta_{ab},\qquad
\int\Xi_a^\dagger\Xi_b\,d^3r=\frac{m}{\mu}\,\delta_{ab}.
\end{equation}
In a fixed-$p_z$ fiber these are transverse integrals, with the common longitudinal normalization removed.
The first-order problem on the full chirality-degenerate manifold is $Wc=\delta E\,Jc$, with
\begin{equation}\label{eq:secular}
\begin{gathered}
J={\rm diag}(I_d,-I_d),\qquad
W=\begin{pmatrix}V&-iC\\ iC&-V\end{pmatrix},\\
V_{ab}=\langle f_a|qA_0|f_b\rangle,\qquad
C_{ab}=\frac{gq}{2m}\int\Xi_a^\dagger(\vS\!\cdot\!\vE)\Xi_b\,d^3r
       =\frac{g}{2\mu}\langle f_a|q\vS\!\cdot\!\vE|f_b\rangle .
\end{gathered}
\end{equation}
Both $V$ and $C$ are Hermitian, but the energy shifts are not the ordinary eigenvalues
of $W$: the signature matrix $J$ must also be included. Since $J^2=I$, the generalized
problem is equivalent to diagonalizing $JW$. A constant unitary change of basis makes
this matrix block diagonal, with blocks $V+iC$ and $V-iC$, whose spectra are complex
conjugates. That change of basis is the return to the chirality basis. Since
$H={\rm diag}(H_L,H_R)$ for arbitrary fields, $V-iC$ is the first-order problem of $H_L$ on
its own unperturbed eigenspace, formed with the left eigenvectors $\tau_3u_a$, and $V+iC$ is
that of $H_R$. Within one block the unperturbed states all carry positive $\tau_3$-charge;
the electric field acts there by breaking the $\tau_3$-pseudo-Hermiticity of the block,
\eq{eq:pseudo}, and this is why the projected matrix is not Hermitian. The two descriptions
are equivalent: the opposite-signature form keeps the conserved metric $\tau_5$ in view, and
the single-block form is the more economical for calculation.

This reduction shows precisely when a diagonal expectation-value argument is sufficient.
If the full projected $C$ vanishes, the shifts are simply the real eigenvalues of $V$.
For a closed single chirality pair, $V=v$ and $C=c$ are numbers and the shifts are $v\pm ic$.
For a larger manifold, diagonal expectation values alone are insufficient: zero diagonal entries do not imply $C=0$,
and a nonzero $C$ need not make the spectrum nonreal. For example,
$V=a\sigma_3$, $C=b\sigma_1$ gives eigenvalues $\pm\sqrt{a^2-b^2}$.
The complete projected matrices must be diagonalized.

The spin-$\frac12$ limit shows how the reality of the Dirac spectrum is realized in this
formalism. At $j=\frac12$, $g=2$, the magnetic operator is a square, $K_B=D^2$ with
$D=\vec\sigma\!\cdot\!\vpi$, and for a static potential the spin--electric term is a commutator,
$q\,\vec\sigma\!\cdot\!\vE=-i[D,qA_0]$. Choosing the basis $\{f_a\}$ of the unperturbed eigenspace
to diagonalize $D$, with $Df_a=k_af_a$, $k_a=\pm k$ and $k^2=\mu^2-m^2$, one finds
\[
C_{ab}=-\frac{i}{2\mu}\,(k_a-k_b)\,V_{ab}.
\]
Hence $C$ vanishes between states with the same sign of $k_a$, and in particular $C_{aa}=0$:
a nondegenerate state has a real first-order shift. Between the two signs, the off-diagonal
blocks of $V-iC$ are those of $V$ multiplied by $1\mp k/\mu$. A similarity transformation by
the positive diagonal matrix ${\rm diag}(t,1)$, with $t^2=(\mu-k)/(\mu+k)$, then turns $V-iC$
into the Hermitian matrix obtained from $V$ by multiplying its off-diagonal blocks by
$\sqrt{1-k^2/\mu^2}=m/\mu$. The first-order shifts are therefore real for every static magnetic
background and potential. Beyond first order the same conclusion follows from the
factorization, for domains on which the corresponding Dirac Hamiltonians are self-adjoint:
each solution splits into mass-$m$ and mass-$-m$ Dirac solutions, which excludes normalizable
complex-energy eigenstates of the squared equation. No analogous commutator identity is
available for $j\ge1$, or for $j=\frac12$ with $g\neq2$.
The spin--electric operator mixes the two $\rho_1$ sectors; it does not preserve a reduction that simply removes
one of them. In intermediate-state sums the norm sign is $\eta s$, so changing $\eta$ reverses the norm sign only
when the frequency sign $s$ is held fixed.

Parity provides another useful simplification, although only under a different choice
of reference problem. For a fixed parity-even potential retained in $H_0$, with the
spin--electric term switched on as a separate
perturbation, the projected $C$ vanishes on an isolated eigenspace containing only one spatial parity.
In particular, $\vS\!\cdot\!\nabla V$ is odd for a central Coulomb potential and connects
$l$ to $l\pm1$, whose energies are not degenerate in the point-Coulomb spinless Klein--Gordon reference problem.
The first-order spin--electric contribution is then absent. Second order involves the opposite chirality-exchange
sector, with intermediate norm signs $\eta s$ of either sign, as in the mixed-norm perturbation theory of
Ref.~\cite{TST2025}. The full coupled-channel Coulomb spectrum and the validity of a finite-coupling continuation
are not established by this selection rule.

\subsection{Instability under a bounded electrostatic perturbation}\label{sec:bounded}

To demonstrate that the mixed-norm secular problem can produce a genuine spectral
instability within the model, we consider an electric perturbation that remains bounded
on the full transverse plane. We take a uniform magnetic field in the symmetric gauge
$\vA=\frac B2(-y,x,0)$, put $b=qB>0$, and fix $p_z=0$. The axisymmetric
electrostatic potential and its associated field are chosen as
\begin{equation}\label{eq:VR}
\begin{aligned}
qA_0=V_R(r_\perp)&=\tfrac12\kappa r_\perp^2e^{-r_\perp^2/R^2},\\
q\vE=-\nabla V_R&=-\kappa\vec r_\perp e^{-r_\perp^2/R^2}
                    \left(1-\frac{r_\perp^2}{R^2}\right).
\end{aligned}
\end{equation}
Here $R>0$ is fixed, $[\kappa]={\rm mass}^3$, and the harmonic limit is $R\to\infty$.
For every finite $R$, both $V_R$ and its gradient are bounded.
This boundedness is important for the spectral argument below: the potential and its
associated spin--electric term give a bounded perturbation of the FV operator.
Since $p_z$ and $J_z=L_z+S_z$ are conserved, we work in a fixed $(p_z,J_z)$ fiber.
In the stable magnetic regime, each nonzero unperturbed eigenvalue in that fiber is isolated and has finite multiplicity.

Before evaluating matrix elements, we identify the degeneracies that the perturbation
can connect. The unperturbed levels are $E^2=m^2+b(2n+1-gm_s)$, where $n$ is the
Landau index.
Different spin projections can coincide only when $g\Delta m_s\in2\mathbb Z$.
The transverse spin--electric operator connects $\Delta m_s=\pm1$, so the corresponding degeneracies require
$g\in2\mathbb Z$. They occur at $g=2$, where $E^2=m^2+2bN$ and
$n=N-\frac12+m_s$, but not at $g=1/j$ for $j\ge1$.
For that latter benchmark, $C$ vanishes within every magnetic eigenspace for the perturbation~\eqref{eq:VR};
its first-order shifts are real. This is not an all-orders stability statement, and it is specific to the
uniform magnetic field, whose degenerate levels at $g=1/j$ differ by $\Delta m_s=2j\ge2$. For $j\ge1$ there is
no counterpart of the commutator identity of \sect{sec:electric}, so nothing protects $C$ in general. In an
inhomogeneous magnetic background, a nondegenerate state with
$C_{aa}=\frac{g}{2\mu}\int qA_0\,\nabla\!\cdot\!(f_a^\dagger\vS f_a)\,d^3r\neq0$ would acquire the nonreal
first-order shift $v\pm ic$ at any nonzero $g$, unless a symmetry of the kind identified below forbids it.

The simplest nontrivial example is the spin-$\frac32$ doublet. For $j=\frac32$,
$g=2$, and $J_z=\frac12$, the $E=m$ two-state manifold consists of
\[
(n,m_\ell,m_s)=\left(0,0,\tfrac12\right),\qquad
(n,m_\ell,m_s)=\left(1,-1,\tfrac32\right).
\]
The lower single-state branch $n=0,m_s=\frac32$, with $E^2=m^2-2b$, belongs to other $J_z$ fibers;
it is not present in this $J_z=\frac12$ fiber. To express the projected matrix
elements compactly, we introduce the dimensionless range parameter and the associated
function
\begin{equation}\label{eq:rangefactor}
a=\frac{2}{bR^2},\qquad F(a)=\frac{1-a}{(1+a)^3}.
\end{equation}
Evaluating the potential and spin--electric matrix elements between the two Landau
states then gives
\begin{equation}\label{eq:boundedmatrix}
V_0=\frac{\kappa}{b(1+a)^2},\qquad
V_1=\frac{2\kappa}{b(1+a)^3},\qquad
|C_{01}|=\frac{|\kappa|}{m}\sqrt{\frac{3}{2b}}\,|F(a)|.
\end{equation}
Thus, with $\bar v=(V_0+V_1)/2$, the first-order shifts are
\begin{equation}\label{eq:krein}
\delta E_\pm^{(1)}=\bar v\pm\frac{\kappa F(a)}{2b}
                 \sqrt{1-\frac{6b}{m^2}}.
\end{equation}
The discriminant changes sign at $b=m^2/6$ whenever $F(a)\neq0$.
At the exceptional range $a=1$, equivalently $bR^2=2$, the first-order splitting vanishes;
no conclusion about higher-order stability follows from~\eqref{eq:krein} at that range.

The projected discriminant identifies a possible instability, but a statement about
eigenvalues of the differential operator also requires control of the perturbation.
For this purpose, we fix
\[
\frac{m^2}{6}<b<\frac{m^2}{2},\qquad R>0,\qquad a\neq1.
\]
Let $K_B$ be the self-adjoint magnetic operator in the fixed $p_z=0$, $J_z=\frac12$ fiber, with domain
$\mathcal D(K_B)$. The unperturbed block $H_L(0)=\tp\otimes K_B/2m+m\tau_3$ is not closed on
$\mathcal D(K_B)\oplus\mathcal D(K_B)$, because $K_B$ acts only on the combination $\varphi+\chi$.
Its closure has the domain
\[
\mathcal D=\bigl\{(\varphi,\chi)\in L^2\oplus L^2:\ \varphi+\chi\in\mathcal D(K_B)\bigr\},
\]
which is $\mathcal D(K_B)\oplus L^2$ in the variables $X=\varphi+\chi$, $Y=\varphi-\chi$. In these variables the
block is $A=\bigl(\begin{smallmatrix}0&m\\ (K_B+m^2)/m&0\end{smallmatrix}\bigr)$, with resolvent
\[
(z-A)^{-1}=\begin{pmatrix}z&m\\ (K_B+m^2)/m&z\end{pmatrix}\bigl(z^2-m^2-K_B\bigr)^{-1},
\]
whose entries are bounded for $z^2-m^2$ outside the spectrum of $K_B$, since
$(K_B+m^2)(z^2-m^2-K_B)^{-1}=-1+z^2(z^2-m^2-K_B)^{-1}$. The operator is thus closed on $\mathcal D$ with nonempty
resolvent set, and this domain is kept fixed.
For finite $R$, $H(\kappa)=H(0)+\kappa\mathcal B_R$ with $\mathcal B_R$ bounded on $L^2\oplus L^2$ is an analytic family.
In each invariant chiral block, $E=m$ is an isolated semisimple eigenvalue of multiplicity two.
Separating the perturbation amplitude from the first-order coefficients makes the
corresponding expansion explicit:
\begin{equation}\label{eq:lambdacoef}
\lambda_\pm=
\frac{(3+a)\pm(1-a)\sqrt{1-6b/m^2}}{2b(1+a)^3},
\qquad
E_\pm(\kappa)=m+\kappa\lambda_\pm+O(\kappa^2).
\end{equation}
Since these coefficients are distinct and nonreal, analytic perturbation theory for an isolated finite
spectral group gives analytic branches with the displayed expansion~\cite[Chap.~II, Sec.~2.3; Chap.~VII, Secs.~1.3 and~2.1]{Kato1980}.
Consequently the eigenvalues are nonreal for all sufficiently small nonzero real $\kappa$.
The permitted smallness depends on $b$ and $R$; no estimate uniform near $a=1$ or $b=m^2/6$ is asserted.

The reality of the shifts below $b=m^2/6$, and the way they leave the real axis above it, reflect a symmetry of
the background. A chiral block is not $\tau_3$-pseudo-Hermitian when $\vE\neq0$, so \eq{eq:pseudo} alone does
not protect its spectrum. Let $\mathcal R_y$ be the reflection $y\to-y$ and $\mathcal K$ complex conjugation in
the standard basis, in which $S_x$ and $S_z$ are real. The antiunitary operator
\[
\Theta=e^{i\pi S_z}\,\mathcal R_y\,\mathcal K,\qquad\Theta^2=1,
\]
sends $(\pi_x,\pi_y)\to(-\pi_x,\pi_y)$, $(S_x,S_y,S_z)\to(-S_x,S_y,S_z)$ and $i\to-i$, and for the axisymmetric
field \eqref{eq:VR} it reverses $E_y$. It therefore leaves $\vpi^{\,2}$, $S_zB$, $i\,\vS\!\cdot\!\vE$ and $V_R$
separately invariant, and it commutes with $J_z$ and with each block $H_L$, $H_R$ for every $\kappa$.
Consequently the spectrum of each chiral block in a fixed $J_z$ fiber is symmetric under complex conjugation.
In a basis of $\Theta$-invariant states the projected matrix $V-iC$ is real, because $V$ is real symmetric and
$C$ is imaginary antisymmetric; in particular $C_{aa}=0$. A nondegenerate level therefore cannot leave the real
axis at first order, and on the doublet the two shifts are either both real or a conjugate pair, as in
\eq{eq:krein}. The value $b=m^2/6$ is where this antiunitary symmetry is broken on the doublet: there the two
eigenvectors of the projected matrix coalesce, an exceptional point of the first-order problem inside each
chiral block, of the kind familiar from pseudo-Hermitian and $\mathcal{PT}$-symmetric
systems~\cite{Mostafazadeh2002,Mostafazadeh2010,Heiss2012}. It follows that both branches $E_\pm(\kappa)$
belong to each chiral block. The relation $\tau_3H_R^\dagger\tau_3=H_L$ maps the spectrum of one block onto the
complex conjugate of the other, which here is the same set, so in the sixteen-component operator every nonreal
eigenvalue of this example is doubly degenerate.

The same calculation also clarifies which parts of the mechanism depend on the spin.
At $g=2$, the lowest two-state manifold for general spin $j\ge1$ consists of
$(n,m_\ell,m_s)=(1,-1,j)$ and $(0,0,j-1)$ in the $J_z=j-1$ fiber, with
$\mu_j^2=m^2+(3-2j)b$. The dimensionless discriminant is
\[
1-\frac{4jb}{\mu_j^2}
=\frac{m^2-3(2j-1)b}{m^2+(3-2j)b}.
\]
For $F(a)\neq0$, the first-order threshold is therefore
$b=m^2/[3(2j-1)]$, one third of the magnetic threshold~\eqref{eq:thr}.
The bounded-perturbation conclusion applies with fixed $b$ between these two thresholds.
For $j=\frac12$, $g=2$, the first-order spectra are instead real, by the commutator identity of
\sect{sec:electric}.
In the harmonic limit this is explicit: apart from the zero-mode singlet, a paired manifold
$(n,m_s)=(N,\frac12),(N-1,-\frac12)$ has shifts
$\frac12(v_1+v_2)\pm\frac{\kappa}{2b}\frac{m}{\mu}$ at $p_z=0$,
with positive discriminant $m^2/\mu^2$; the factor $m/\mu$ is the one derived there.

For comparison with the bounded example, it is useful to retain the harmonic limit
as a projected first-order calculation. Higher Landau manifolds can then be treated
using the same secular construction.
On a fixed $(N,J_z)$ manifold, $V$ is diagonal and $C$ is tridiagonal:
\begin{equation}\label{eq:harmonicmatrix}
\begin{aligned}
V_{m_s}&=\frac{\kappa}{b}(2n+m_\ell+1),\qquad m_\ell=J_z-m_s,\\
|C_{m_s+1,m_s}|&=\frac{|g\kappa|}{2\sqrt2\,\mu\sqrt b}\sqrt{n+1}
                    \sqrt{j(j+1)-m_s(m_s+1)}.
\end{aligned}
\end{equation}
Only allowed Landau states, $n\ge\max(0,-m_\ell)$, are included.
In units of $\kappa/b$, the shifts are a common constant plus the eigenvalues of
${\rm diag}(m_s)\mp i\beta T$, where $\beta=g\sqrt b/(2\sqrt2\mu)$ and $T$ contains the square roots in~\eqref{eq:harmonicmatrix}.
For $j=\frac32$, and fibers containing the full indicated manifold, the first-collision thresholds are
$b/m^2=1/6$ ($N=0$), $0.1445$ ($N=1$), $0.1317$ ($N=2$), $0.0940$ ($N=10$), and $0.0482$ ($N=100$).
For $N=1$ the value is the positive root of
$162x^3-81x^2+36x-4=0$, with $x=b/m^2$.
The decrease with $N$ does not establish an instability of the full operator at arbitrary field strength.
The harmonic perturbation is unbounded, its matrix elements grow as $\kappa N/b$, and the level spacing shrinks
as $b/\mu_N$; the condition $|\kappa|N\mu_N/b^2\ll1$ fails at fixed $\kappa$ as $N$ grows.
A bounded global completion also changes the high-$N$ matrices. We make no uniform high-level spectral claim.

Finally, we compare the analytic two-state results with finite-basis spectra of the
full FV Hamiltonian
(Appendix~\ref{app:num}). At $b=0.3m^2$, $\kappa=10^{-3}m^3$, and $J_z=\frac12$, the harmonic calculation gives
$E/m=1.004959\pm0.001480\,i$, unchanged to the displayed precision over the tested cutoffs.
The first-order imaginary magnitude is $1.491\times10^{-3}m$, and the numerical-to-first-order ratio approaches
one as $\kappa$ decreases. The bounded wells at $a=0.25$ and $0.5$ give imaginary magnitudes
$5.76\times10^{-4}m$ and $2.22\times10^{-4}m$, compared with $5.72\times10^{-4}m$ and $2.21\times10^{-4}m$
at first order. Low-lying spin-$\frac12$ test states have real eigenvalues within numerical precision.
These checks support the analytic bounded-background mechanism; they are not exact infinite-dimensional diagonalizations.
In particular, at $g=2$ and $j\ge1$, the specified bounded perturbation can make a two-state magnetic manifold complex
while the unperturbed magnetic spectrum remains real. At $g=1/j$, no imaginary first-order correction occurs for this perturbation in a uniform magnetic field.

\subsection{Nonrelativistic limit and the meaning of \texorpdfstring{$g$}{g}}\label{sec:nr}

The magnetic interpretation of $g$ can be read directly from the low-momentum limit.
For a channel in which $K$ is a $c$-number, the positive eigenvalue of \eq{eq:HFV} is
$\sqrt{m^2+K}=m+\frac{K}{2m}-\frac{K^2}{8m^3}+\dots$, so
\begin{equation}\label{eq:NR}
H_{\rm NR}=m+\frac{\vpi^{\,2}}{2m}-\frac{g\,q}{2m}\,\vS\!\cdot\!\vB\mp\frac{i\,g\,q}{2m}\,\vS\!\cdot\!\vE+qA_0+\dots .
\end{equation}
The magnetic term identifies the moment as $\vec\mu=g\,\frac{q}{2m}\vS$, and hence
fixes the meaning of the parameter $g$ used throughout the paper. This identification
is the only purpose for which we use \eq{eq:NR}. When $[K,qA_0]\neq0$ the Feshbach--Villars reduction generates, as for Klein--Gordon, commutator terms
beyond the order displayed, and \eq{eq:NR} is not the complete nonrelativistic Hamiltonian of the interacting problem.
The imaginary electric term is the spin-$j$ image of the $\vec\sigma\!\cdot\!\vE$ term whose role in the eight-component
theory was clarified in Ref.~\cite{TST2025}: for a static potential it reads $\pm\frac{ig}{2m}\vS\!\cdot\!\nabla V$, it must
accompany every change of the potential, and its first-order contribution must be combined with the potential in \eqref{eq:secular}.

\section{Hyperbolicity}\label{sec:hyp}

The spectral questions considered above are separate from the propagation of wave
fronts. To examine the latter, it is most convenient to return to the covariant
second-order equation. In \eq{eq:2nd}, the potential enters through $D_\mu D^\mu$
at first-derivative order and below, while the field strength
enters algebraically, at derivative order zero. The principal symbol of the covariant equation is therefore
$k^2\,\mathbf 1_{2j+1}$, independently of $g$. The equation is normally
hyperbolic and its characteristic surfaces are null for smooth prescribed fields of
any strength. Local Cauchy well-posedness is understood with the usual regularity and
domain assumptions. Thus the model does not exhibit Velo--Zwanziger acausality, as was
already noted for this equation in Ref.~\cite{Kruglov2001} by the method of
Ref.~\cite{VeloZwanziger1969a}. This is a
property of the representation rather than a repair of the Rarita--Schwinger theory: the
vector-spinor description retains its constraints, and with them the acausal characteristics
of Refs.~\cite{VeloZwanziger1969a,VeloZwanziger1969b}. The Feshbach--Villars system \eqref{eq:HFV} is a first-order-in-time rewriting of
mixed differential order (its highest spatial derivatives sit in a nilpotent block); its Cauchy problem is well posed
exactly when that of \eq{eq:2nd} is, and the characteristic analysis is to be done on the second-order equation.

This conclusion should be distinguished from the corresponding result in the
vector-spinor theory. There, the minimally gauged projector
equation has $g=\frac13$ [Eq.~(118) of Ref.~\cite{NKR2006}], below even Belinfante's $1/s=\frac23$ of the minimally
coupled RS equation, and is acausal; causality holds only for $g\in\{0,2\}$~\cite{NKR2006}, because the constraint
elimination feeds $F_{\mu\nu}$ into the first-derivative terms. In the chiral theory causality is blind to $g$. The same
is seen in the $(\frac32,0)\oplus(0,\frac32)$ tensor-spinor construction of
Ref.~\cite{Acosta2015}, whose characteristic determinant is $(n^2)^8$ at $g=\frac23$.
In both cases the algebraic Pauli term leaves the principal symbol unchanged. By contrast,
Guertin and Wilson~\cite{GuertinWilson1977} showed that even spin-$0$ Hamiltonian theories propagate noncausally once
external-field couplings enter the principal part of the equation; none does here.

\section{Uniform magnetic field: exact Landau levels and the exceptional point}\label{sec:landau}

\subsection{Landau spectrum and stability boundary}\label{sec:landauspectrum}

The uniform magnetic background allows the channel structure used above to be solved
exactly. We take $\vB=B\hat z$, $\vE=0$, and the Landau gauge
$\vA=(0,Bx,0)$. The transverse kinetic operator has eigenvalues
$|qB|(2n+1)$, with Landau index $n=0,1,2,\dots$, while $\vS\!\cdot\!\vB$ has eigenvalues $Bm_s$.
Since these operators commute and $H_L=H_R$, the Hamiltonian separates into channels
labeled by $(n,p_z,m_s)$. In each channel, $K$ is the number
\begin{equation}\label{eq:Kland}
K=p_z^2+|qB|(2n+1)-g\,qB\,m_s ,
\end{equation}
so that, by the identity below \eq{eq:free},
\begin{equation}\label{eq:landau}
E^2=m^2+p_z^2+|qB|(2n+1)-g\,qB\,m_s .
\end{equation}
For $j=\frac12$, $g=2$ this is the Dirac spectrum, $E^2=m^2+p_z^2+2|qB|N$ with $N=n+\frac12-{\rm sgn}(qB)m_s$, including the lowest branch $E^2=m^2+p_z^2$; for $j=1$, $g=2$ it coincides with the spectrum of the $W$ boson~\cite{NielsenOlesen1978}. We have
confirmed \eq{eq:landau} by diagonalizing the full $16\times16$ matrix for $j=\frac32$ numerically, for either sign of $g$;
the spectrum agrees to machine precision, and the matrix is $\tau_5$-pseudo-Hermitian throughout. Each level appears once in
each chirality reduction of \sect{sec:states}. The spectrum \eqref{eq:landau} is consistent with the
constant-field solutions of Ref.~\cite{Kruglov2001}, whose quantization condition [Eq.~(66) there]
reduces to it in a purely magnetic field, up to the labeling of the spin projection.

\begin{figure}[tb]
\centering
\includegraphics[width=\textwidth]{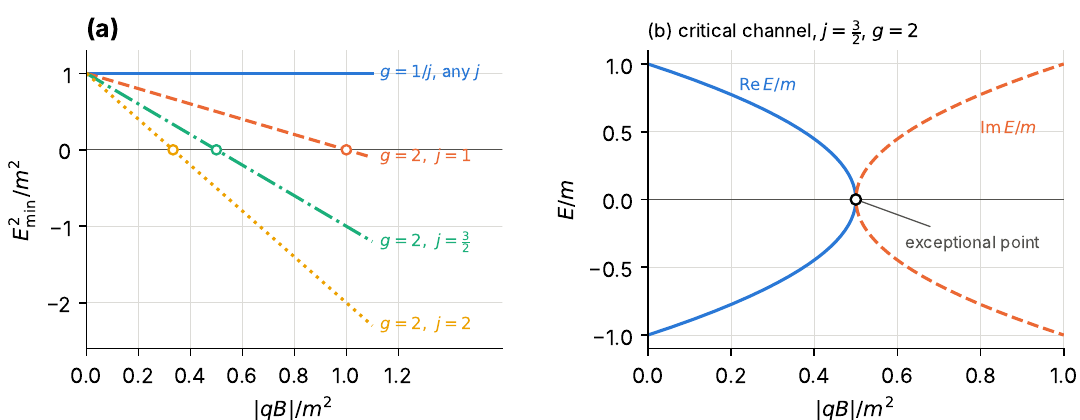}
\caption{(a) Lowest Landau level, \eq{eq:Emin}: at $g=1/j$ it is $m^2$ for every spin; at $g=2$,
$E^2_{\min}=m^2-(2j-1)|qB|$, with circles at $|qB|_c=m^2/(2j-1)$. (b) The two eigenvalues
$\pm\sqrt{m^2-2|qB|}$ of the critical channel ($n=0$, $p_z=0$, $m_s=\frac32$) of the $j=\frac32$, $g=2$ Feshbach--Villars block: real below and
imaginary, $\pm i\sqrt{2|qB|-m^2}$, above the exceptional point at $|qB|=m^2/2$.}
\label{fig:landau}
\end{figure}

To determine the stability boundary, we minimize the squared frequency over all
orbital and spin channels. The minimum occurs at $n=0$, $p_z=0$, and
$m_s=j\,{\rm sgn}(g\,qB)$, giving
\begin{equation}\label{eq:Emin}
E^2_{\min}=m^2+|qB|\,(1-|g|\,j),\qquad |qB|_c=\frac{m^2}{|g|\,j-1}\quad(|g|\,j>1).
\end{equation}
It follows that all channel frequencies remain real at every field strength if and only
if $|g|\le1/j$. Belinfante's benchmark $g=1/j$ lies on this boundary: its lowest
positive-frequency level stays at $E=m$ for every field strength, generalizing the
Dirac lowest Landau branch to spin $j$. In particular, $g=\frac23$, also obtained
in Ref.~\cite{Acosta2015}, has this property within the present model. The benchmark
$g=2$ instead lies in the unstable region for every $j\ge1$, with critical fields
\begin{equation}\label{eq:thr}
|qB|_c\big|_{g=2}=\frac{m^2}{2j-1}=\infty,\ m^2,\ \tfrac12m^2,\ \tfrac13m^2\qquad\text{for }j=\tfrac12,1,\tfrac32,2 ,
\end{equation}
as summarized in \tabl{tab:thr} and \fig{fig:landau}(a). The $j=1$ entry is the Nielsen--Olesen instability~\cite{NielsenOlesen1978}.
The formula \eqref{eq:Emin} is set by $g$ and $j$ alone; to linear order in the field it is also what the effective-theory
treatment of the vector-spinor spin-$\frac32$ sector gives~\cite{Benakli2025}. In the present model, however, \eq{eq:landau} is exact. Its zero is therefore an exact
threshold of the specified equation rather than an extrapolation of a weak-field
expansion. The condition $|g|j>1$ for its existence is the known condition for pair
production in a purely magnetic field~\cite{MarinovPopov1972,Kruglov2001}; the unstable
channel is its counterpart in the wave equation. What is examined next, and is not contained
in those works, is how this threshold appears in the FV evolution operator.

\begin{table}[tb]
\centering
\small
\setlength{\tabcolsep}{4pt}
\caption{Lowest Landau level of the chiral second-order model, \eq{eq:Emin}, at the two benchmark values of $g$.
At $g=1/j$ the lowest level is $E=m$ for every $j$ and every field.}
\label{tab:thr}
\begin{tabular}{cllll}
\toprule
$j$ & $E^2_{\min}$ at $g=2$ & $|qB|_c$ at $g=2$ & $E^2_{\min}$ at $g=1/j$ & \\
\midrule
$\frac12$ & $m^2$ & $\infty$ & $m^2$ & Dirac zero mode \\
$1$ & $m^2-|qB|$ & $m^2$ & $m^2$ & Nielsen--Olesen~\cite{NielsenOlesen1978} \\
$\frac32$ & $m^2-2|qB|$ & $m^2/2$ & $m^2$ & $gs>1$: Refs.~\cite{MarinovPopov1972,Kruglov2001}; $g=\frac23$: \cite{Acosta2015} \\
$2$ & $m^2-3|qB|$ & $m^2/3$ & $m^2$ & \\
\bottomrule
\end{tabular}
\end{table}

\subsection{Exceptional-point structure and time evolution}\label{sec:exceptional}

For definiteness, let us take $gqB>0$, so that the critical spin projection is
$m_s=j$. The FV block in this channel is $h(K)=\tp\frac{K}{2m}+m\tau_3$,
whose eigenvalues are
$\pm\sqrt{m^2+K}$. At $K=-m^2$,
\begin{equation}\label{eq:EP}
h(-m^2)=\frac m2\begin{pmatrix}1&-1\\1&-1\end{pmatrix},\qquad h(-m^2)^2=0,\qquad{\rm rank}\,h(-m^2)=1 .
\end{equation}
These identities show that the critical block is a $2\times2$ Jordan block $J_2(0)$.
Its eigenvector $u=(1,1)^T$ is neutral, $u^\dagger\tau_3u=0$, and
$v=\frac1m(1,-1)^T$ completes the chain, $h\,v=u$, $h\,u=0$. The block occurs once per chirality, so in the full
sixteen-component problem the critical channel is $J_2(0)\oplus J_2(0)$, algebraic multiplicity four and geometric
multiplicity two; in either chirality reduction of \sect{sec:states} it is a single $J_2(0)$ with the $\tau_5$-neutral
eigenvector $(u,\pm u)\otimes|m_s{=}j\rangle$. This is a statement about the fixed channel $(n{=}0,p_z{=}0,m_s{=}j)$;
in infinite space $p_z$ is continuous and the Landau levels degenerate, so it is an exceptional point of the channel
Hamiltonian, not an isolated one of the full differential operator. Its time-domain meaning is explicit: at threshold
$e^{-ith}=1-ith$, linear growth for initial data along $v$; above it the pair is $\pm i\gamma$ with
\begin{equation}\label{eq:growth}
\gamma(p_z)=\sqrt{\,|qB|\,(|g|j-1)-m^2-p_z^2\,}
\end{equation}
wherever the radicand is positive, and the eigenvectors are again neutral. This is the Krein-space mechanism described in
general terms in Refs.~\cite{Bognar1974,Mostafazadeh2002,Mostafazadeh2010} and invoked, without an example, in the
perturbation theory of the eight-component equation~\cite{TST2025}: a pair of real eigenvalues of a pseudo-Hermitian
operator becomes complex by colliding at an exceptional point~\cite{Heiss2012} at which the eigenvectors become neutral.

\subsection{Positive conserved forms}\label{sec:positiveform}

The significance of the exceptional point can also be expressed without choosing a
particular normalization of its eigenvectors. We ask whether the channel generator
admits a time-independent positive-definite conserved quadratic form. With
$\omega^2=m^2+K$ and the variables $X=\varphi+\chi=\Xi$, $Y=\varphi-\chi=\pi_0\Xi/m$, the channel equation is
\begin{equation}\label{eq:XY}
\begin{gathered}
i\partial_t\begin{pmatrix}X\\Y\end{pmatrix}=A_\omega\begin{pmatrix}X\\Y\end{pmatrix},\qquad
A_\omega=\begin{pmatrix}0&m\\ \omega^2/m&0\end{pmatrix},\\
A_\omega^\dagger\,\eta_\omega=\eta_\omega A_\omega,\quad \eta_\omega={\rm diag}(\omega^2/m^2,\,1).
\end{gathered}
\end{equation}
The constant Hermitian forms with $A_\omega^\dagger\eta=\eta A_\omega$ are exactly
$\eta=\bigl(\begin{smallmatrix}d\,\omega^2/m^2&b\\ b&d\end{smallmatrix}\bigr)$, $b,d$ real, positive definite iff $d>0$ and
$d^2\omega^2/m^2>b^2$. Such positive-definite forms exist if and only if $\omega^2>0$.
For $\omega^2>0$ the representative $\eta_\omega$ shows that the stable
channel conserves the positive quantity $\omega^2|\Xi|^2/m^2+|\pi_0\Xi|^2/m^2$, proportional to the Klein--Gordon energy of the channel, in
addition to the indefinite charge $\varphi^\dagger\varphi-\chi^\dagger\chi={\rm Re}(X^\dagger Y)$. At $\omega^2=0$ the positive representative degenerates, and no positive-definite form can make a non-trivial Jordan block self-adjoint; for $\omega^2<0$ the
non-real eigenvalues exclude one as well. The exceptional point is therefore the boundary beyond which the channel generator
admits no time-independent positive-definite conserved form. The mechanism is
generic to any Feshbach--Villars channel whose effective frequency squared crosses zero; what spin $\frac32$ supplies is the
channel in which this happens at $|qB|=m^2/2$ for $g=2$, and never for $|g|\le\frac23$. It cannot occur for $j=\frac12$ with
$g=2$, where $K\ge0$; it first appears at $j=1$.

\section{Comparison: Rarita--Schwinger, Nielsen--Olesen, and the role of \texorpdfstring{$g$}{g}}\label{sec:compare}

The preceding results involve four logically distinct questions: the characteristics,
the well-posedness of the Cauchy problem, the reality of the spectrum, and the positivity
of the norm. Keeping these questions separate is essential when comparing the chiral
model with other descriptions of charged higher-spin particles. For the minimally coupled Rarita--Schwinger equation in a uniform magnetic
field, with $\beta=2|eB|/3m^2$, the characteristic speed transverse to the field is
$v_\perp=(1-\beta^2)^{-1/2}>1$ for $0<\beta<1$~\cite{VeloZwanziger1969a}: wave fronts leave the light cone for
arbitrarily weak fields. For a timelike normal the characteristic determinant is
$n_0^{16}(1-\beta^2)^4$~\cite{VeloZwanziger1969a,SiahaanThesis}, so hyperbolicity itself is lost at $\beta=1$, i.e.\
$|eB|=\frac32m^2$, a separate event; and Seetharaman, Prabhakaran and Mathews~\cite{Seetharaman1975} found that at the
same field the Landau energies become complex and the norm, positive definite for $\beta<1$, becomes indefinite. In
\eq{eq:2nd} the first two phenomena do not occur at all: the characteristics are null for every field (\sect{sec:hyp}).
For $|g|j>1$, the magnetic frequencies instead cease to be real at $|qB|_c=m^2/(|g|j-1)$, through the channel exceptional
point of \sect{sec:landau}. The metric is indefinite throughout; what changes at this threshold is the spectral decomposition:
the colliding eigenvectors become neutral, and the real-frequency separation into positive- and negative-norm states fails.
A complex frequency is not by itself a failure of hyperbolicity: \eq{eq:2nd} above threshold is a hyperbolic equation with
a negative effective mass squared in one channel, exactly as the $W$ field above $eB=m_W^2$~\cite{NielsenOlesen1978} and
the charged higher-spin string states in strong fields~\cite{FerraraPorratiTelegdi1992,PorratiRahmanSagnotti2011}. The two
thresholds, $m^2/2$ here and $\frac32m^2$ for Rarita--Schwinger, are of the same order; they differ in kind.

The magnetic calculation therefore establishes an exponentially growing mode of the
classical chiral wave equation in a prescribed background, while the characteristics
remain lightlike. It does not establish a vacuum decay
rate, a positive-norm fermionic quantization, or a stable end point. For the same equation,
pair-production probabilities and the one-loop effective Lagrangian in constant fields were
computed in Ref.~\cite{Kruglov2001}, with production in a purely magnetic field for $gs>1$
in agreement with Ref.~\cite{MarinovPopov1972}; how those field-theoretic results are to be
related to the indefinite-metric one-particle structure studied here is not addressed in this paper. For the $W$ boson the end point is a condensate; a
fermionic field cannot simply inherit that interpretation, and for an indefinite-metric one-particle equation the
quantum meaning of a tachyonic channel is open. The comparison with Nielsen--Olesen holds at the level of the linear
spectral mechanism and no further.

The role of the gyromagnetic ratio further illustrates why these criteria should not
be conflated. In the vector-spinor theory causality fixes $g=2$~\cite{NKR2006};
in the chiral model the characteristics are $g$-independent, and the benchmarks have different relations to the magnetic stability
boundary $|g|=1/j$ of \eq{eq:Emin}: Belinfante's value is stable for every spin, $g=2$ only for
$j=\frac12$. For a point-like spin-$\frac12$ particle with an anomalous moment the same formula gives, formally,
$E^2_{\min}=m^2-\frac{g-2}{2}|qB|$ and an instability at $|qB|\simeq2m^2/(g-2)$, approximately $860\,m^2$ when the electron's weak-field anomaly is inserted. This is a formal extrapolation, not a prediction for QED at such fields; for $j\ge1$ at $g=2$ the threshold sits at $m^2/(2j-1)$. Whether such a threshold lies within the regime of validity depends on the
interpretation of \eq{eq:2nd}. As a specified wave equation, it defines an exact model.
If it is instead regarded as the leading term of an effective theory, the omitted
operators must remain small near the threshold. With
$\alpha_g=|g|j-1$ and the illustrative term of \sect{sec:eq}, $E^2_{\min}=m^2-\alpha_g|qB|+c\,(qB)^2/\Lambda^2$, the correction at the
uncorrected threshold $|qB|=m^2/\alpha_g$ is $\delta E^2/m^2=c\,m^2/(\alpha_g^2\Lambda^2)$, so the condition is
$\sqrt{|c|}\,m/(\alpha_g\Lambda)\ll1$---not uniform in $g$: it is $\Lambda\gg m$ for the principal benchmark $j=\frac32$, $g=2$
($\alpha_g=2$), but as $|g|j\to1^+$ the threshold recedes to large fields and no hierarchy in $\Lambda/m$ alone controls it; with
$\Lambda=m$ the zero disappears for $c>\alpha_g^2/4$, i.e.\ $c>1$ at the principal benchmark. The effective-theory analysis of Ref.~\cite{Benakli2025} is perturbative in the
field and does not by itself license the extrapolation to the zero. We therefore make no claim about composite spin-$\frac32$
states.

\section{Conclusions and outlook}\label{sec:outlook}

For the chiral second-order equation of Refs.~\cite{Hurley1971,Hurley1972,Hurley1974,Kruglov2001}, obtained here from the
Poincar\'e-projector construction, the FV rewriting preserves the known lightlike characteristics and exposes two distinct
spectral mechanisms. A purely magnetic channel develops a Jordan block when its effective squared frequency reaches zero;
a bounded electrostatic perturbation can instead make a degenerate doublet complex before that magnetic
threshold is reached, through the breaking of an antiunitary symmetry of each chiral block. The latter statement follows locally in the perturbation amplitude from the nonreal first-order
coefficients in \eq{eq:lambdacoef}, away from the exceptional range $a=1$.
Neither result identifies a physical single-species quantum theory. The following questions remain outside the present scope.

The first unresolved issue is the selection of a physical sector. Section~\ref{sec:states}
shows that a magnetic background admits invariant reductions with a single-species mode
count, without selecting a covariant physical one. Section~\ref{sec:electric} then shows
how an electric perturbation mixes the chirality-exchange sectors, and
Sec.~\ref{sec:bounded} supplies a bounded example with nonreal branches at $g=2$.
For $j=\frac12$, the Dirac relation provides a covariant choice that is preserved by
the interaction. For $j=\frac32$, the customary free Weinberg--Joos relation is cubic
in momentum, and preservation of a gauged version by \eq{eq:2nd} has not been established.
A candidate relation must be gauge covariant, have the correct free limit, and be
compatible with the operator domains as well as the dynamics. In a purely magnetic
frame, $H_L=H_R$, so an operator that merely intertwines the two blocks belongs to
their commutant. Since that commutant contains many operators, intertwining alone does
not identify a physical reduction. Until such a relation is established, the model
must be interpreted as the doubled system, with the mixed-norm perturbation theory
of Ref.~\cite{TST2025} providing the relevant calculus.

A second question concerns the Coulomb problem. This requires a separate coupled-angular-channel
analysis because, at fixed total angular momentum, $\vS\!\cdot\!\hat r$ acts on several
orbital channels. Replacing $\vec\sigma/2$ by $\vS$ does not preserve the spin-$\frac12$
identities used in the Biedenharn reduction of the eight-component theory. It remains
to be determined whether an analogous reduction is useful here and what critical
coupling would follow from the resulting radial equations. The parity selection rule
established above does not answer these questions, and no critical-charge estimate is
made in this work.

Compton scattering offers a further test of the interacting model. Writing
\eq{eq:2nd} as $[\Gamma^{\mu\nu}D_\mu D_\nu+m^2]\Xi=0$, with
$\Gamma^{\mu\nu}=g^{\mu\nu}-i\,g\,S^{\mu\nu}$, gives the propagator and the
one-photon vertex $-(\Gamma^{\nu\mu}p'_\nu+\Gamma^{\mu\nu}p_\nu)$, together with
the seagull tensor $\Gamma^{\mu\nu}+\Gamma^{\nu\mu}=2g^{\mu\nu}$ of the
second-order formalism~\cite{SiahaanThesis}. The Ward identity holds for every $g$,
but the high-energy behavior of amplitudes between specified external states remains
to be examined. Existing results provide benchmarks rather than a conclusion for the
present representation. In the second-order spin-$\frac12$ theory, the total cross
section vanishes at high energy only for $g=2$~\cite{DelgadoAcosta2011}; in the pure
$(\frac32,0)\oplus(0,\frac32)$ tensor-spinor construction at $g=\frac23$, the
forward cross section is energy-independent while the total cross section
grows~\cite{Acosta2015}. In the vector-spinor projector theory at $g=2$, the forward
amplitude is energy-independent~\cite{DelgadoAcostaNapsuciale2009}. A corresponding
calculation for the present model would extend the eight-component spin-$\frac12$
Compton analysis of Ref.~\cite{RobsonSutanto2001b}. It may also show that neither
benchmark value of $g$ is sufficient without additional structure.

\section*{Acknowledgments}
The author is grateful to Dr.\ Paulus Cahyono Tjiang for helpful discussions of the
Feshbach--Villars procedure, and to S.~I.~Kruglov for bringing Ref.~\cite{Kruglov2001} and the
related earlier literature to his attention after the first version of this paper appeared.
This work was supported by LPPM-UNPAR.

\section*{Data availability}
The symbolic and numerical scripts used for the checks described in this work are available from the author on reasonable request.
The analytic expressions and finite-basis implementation details are given in the text and Appendix~\ref{app:num}.

\section*{Use of AI tools}
ChatGPT (OpenAI) and Claude (Anthropic) were used to assist with language editing, manuscript
organization, and selected symbolic and numerical consistency checks. The author reviewed and edited
all such material and takes full responsibility for the content of this article.

\appendix
\section{Numerical details for the bounded-electric example}\label{app:num}

This appendix describes the finite-basis calculations used to check the projected
results of Sec.~\ref{sec:bounded}. The calculations use two-dimensional Cartesian
oscillator states $|N_x,N_y\rangle$ of frequency
$\omega_B=qB/2$ and unit oscillator mass. The operators
$p_\perp^2+\omega_B^2r_\perp^2=qB(N_{\rm tot}+1)$ and $L_z$ preserve
$N_{\rm tot}=N_x+N_y$, so the restriction of
$\vpi_\perp^{\,2}=qB(N_{\rm tot}+1-L_z)$ is exact in the basis $N_{\rm tot}\le N_{\rm keep}$.
The polynomial operators $r_\perp^2$ and $x,y$ change $N_{\rm tot}$ by at most two;
interior matrix elements can therefore be checked independently of the upper-edge truncation.
The gauge is $\vA=\frac B2(-y,x,0)$ and $p_z=0$.
Numerical quantities are nondimensionalized by $m$, with $m=q=1$, $g=2$, and $b/m^2=0.3$ for the tabulated examples.

For the bounded background, functions of $r_\perp^2$ are evaluated by functional calculus on its truncated matrix.
This is a numerical approximation: generally $P_N f(r_\perp^2)P_N\neq f(P_Nr_\perp^2P_N)$.
It is not an exact projection of the continuum Gaussian multiplier, and its convergence is tested by increasing
$N_{\rm keep}$ and comparing projected entries with the analytic formulas in \sect{sec:bounded}.

Once these orbital matrices are assembled, they are combined with the spin and FV
indices to form $H={\rm diag}(H_L,H_R)$, with
\[
H_{L,R}=\tp\otimes\frac{K_{L,R}}{2m}+m\tau_3\otimes I+I\otimes V_R,
\qquad
K_{L,R}=K_B\mp igq\vS\!\cdot\!\vE.
\]
The truncated products representing $q\vS\!\cdot\!\vE$ are symmetrized in the bounded case.
The full dimension is $4(2j+1)\,\#\{(N_x,N_y):N_x+N_y\le N_{\rm keep}\}$.
The reported near-doublet eigenvectors are selected using
$\langle N_{\rm tot}\rangle\le5$ in the ordinary norm as a localization diagnostic,
$\langle J_z\rangle=\frac12$, and a real part within $3|\kappa|/b$ of
$m+\frac12(V_0+V_1)$.
For the projected first-order calculation, $K_B$ is diagonalized first and $J_z$ is diagonalized within each degenerate
subspace; the resulting $V\pm iC$ blocks are compared with the closed forms in \sect{sec:bounded}.

\begin{table}[htbp]
\centering
\small
\setlength{\tabcolsep}{4.2pt}
\caption{Finite-basis eigenvalues near the $J_z=\frac12$ doublet for $j=\frac32$, $g=2$, $p_z=0$, and $b/m^2=0.3$.
The conjugate pair is represented by its common real part and positive imaginary magnitude.
Harmonic entries agree to the displayed precision for $N_{\rm keep}=8,12,16$; bounded entries are the values at
$N_{\rm keep}=12,16$. The ratio compares the numerical imaginary magnitude with its first-order value.}
\label{tab:numerics}
\begin{tabular}{@{}lccccc@{}}
\toprule
Background & $\kappa/m^3$ & $\operatorname{Re}E/m$ & $|\operatorname{Im}E|/m$
 & $|\operatorname{Im}\delta E^{(1)}|/m$ & Ratio\\
\midrule
Harmonic & $2\times10^{-3}$ & $1.009839$ & $0.002941$ & $2.981\times10^{-3}$ & $0.986$\\
Harmonic & $1\times10^{-3}$ & $1.004959$ & $0.001480$ & $1.491\times10^{-3}$ & $0.993$\\
Harmonic & $5\times10^{-4}$ & $1.002490$ & $0.000743$ & $7.454\times10^{-4}$ & $0.997$\\
Bounded, $a=0.25$ & $1\times10^{-3}$ & $1.002771$ & $0.000576$ & $5.724\times10^{-4}$ & $1.006$\\
Bounded, $a=0.5$ & $1\times10^{-3}$ & $1.001729$ & $0.000222$ & $2.208\times10^{-4}$ & $1.005$\\
\bottomrule
\end{tabular}
\end{table}

Table~\ref{tab:numerics} summarizes the comparison with first-order perturbation
theory. By the antiunitary symmetry of \sect{sec:bounded}, each chiral block contributes the same
conjugate pair, so every listed eigenvalue occurs twice in the full matrix. For the three harmonic rows, the predicted centers are, respectively,
$1.010000$, $1.005000$, and $1.002500$ in units of $m$.
For the bounded rows they are $1.002773333$ and $1.001728395$.
The differences between these centers and the numerical real parts are consistent with the expected
$O(\kappa^2)$ correction.
Narrower wells were also run: at $a=2$, for which the well range is comparable to the magnetic length, the
imaginary magnitude still changes between $N_{\rm keep}=12$ and $16$, so that case is not converged in the
tested bases and is not reported.
The spin-$\frac12$, $g=2$ harmonic check at $\kappa/m^3=10^{-3}$ gives real interior eigenvalues to $10^{-15}$
for the tested cutoffs $N_{\rm keep}=8,16$.
These observations establish numerical agreement over the stated truncations, not exact cutoff independence or an
exact diagonalization of the infinite-dimensional harmonic problem.


\end{document}